\documentclass[aps,pra,twocolumn,showpacs]{revtex4}
\usepackage{amsfonts}
\usepackage{graphicx}
\usepackage{amsmath}
\usepackage{amssymb}

\begin{document}

\title{Probing quantum entanglement, quantum discord, classical correlation
and quantum state without disturbing them}
\author{Zhen-ni Li}
\author{Jia-sen Jin}
\author{Chang-shui Yu}
\email{quaninformation@sina.com;ycs@dlut.edu.cn}
\affiliation{Department of Physics, Dalian Technology of University, Dalian 116023, China}

\begin{abstract}
In this paper, we present schemes for a type of one-parameter bipartite
quantum states to probe the quantum entanglement, the quantum discord, the
classical correlation and the quantum state based on the cavity QED. It is
shown that our detection does not influence all these measured quantities.
We also discuss how the spontaneous emission introduced by our probe atom
influences our detection.
\end{abstract}

\maketitle

\section{introduction}

In quantum dynamics, it is important to study whether a given bipartite
quantum state is entangled, separable, quantum correlated, or classically
correlated. Until now many efforts has been taken in this region. As we
know, quantum entanglement is a very useful physical resource in quantum
information processing. In recent years, the quantification of entanglement
has attracted much more attention [1-8], in particular, entanglement of
bipartite pure states and two-qubit mixed states has been obtained good
understanding [1-5,8]. However, quantum entanglement is not the only quantum
correlation in quantum dynamics. Quantum discord [9,10], first introduced by
Ollivier and Zurek [10], can effectively capture the quantum correlation of
quantum system. It has been shown that quantum discord can lead to a speedup
in some quantum information tasks [11-14]. It is especially worth noting
that quantum discord is not consistent with quantum entanglement in general
cases [15]. Quantum discord captures quantum correlation even more general
than entanglement. Specially separable mixed states, which have no
entanglement, have proven to include nonzero quantum discord. Quantum
discord is not always larger than quantum entanglement.

However, most interests of quantifying entanglement and quantum discord are
focused in a pure mathematical frame [16,17,18]. For example, since quantum
entanglement is not an observable in the strict quantum mechanical frame, no
directly measurable observable has been found until now, to describe the
entanglement of a given arbitrary quantum state, owing to the unphysical
quantum operations in the usual entanglement measure, such as the complex
conjugation of concurrence [2] and the partial transpose of negativity [1].
In recent years, some interesting methods have been proposed to construct
direct observables related to entanglement [19-22], which can be used to
precisely measure the entanglement in contrast to entanglement witness [23]
and have less observables compared with quantum state tomography [24,25].
However, they require the simultaneous multiple copies of given quantum
states, which brings new difficulties to the experiment. Quantum discord as
a measure of quantum correlation can only be analytically calculated for
some special states [15,16], unless the invariational definition of quantum
discord is considered [26]. The key difficulty lies in the analytical
calculation of the classical correlation, since quantum discord is defined
as the difference between the total correlation (i.e. quantum mutual
information) and the classical correlation [9,10]. Although the dynamical
behavior and some operational understanding of quantum discord in quantum
state evolution has attracted increasing interests recently [27-29], there
still exist an open question how one can construct several directly
measurable observables related to quantum discord.

In this paper, we propose a scheme to probe the entanglement, discord, and
the classical correlation of a type of one-parameter bipartite quantum
states based on the cavity QED [30-33]. Although the entanglement of
one-parameter quantum state such as Werner state [34], isotropic state and
so on [35,36] can be measured based on simple von Neumann measurements
without the requirement of simultaneous copies of the state, these
measurements usually cover projections on two qubits. However, our
measurement is only performed on the probe qubit. In addition, the
distinguished advantage of our scheme is that one can probe quantum
entanglement, quantum discord and the classical correlation by introducing a
probe qubit to interact with the measured systems, but the entanglement, the
discord and the classical correlation of the system are not disturbed. In
particular, in some cases, one can even realize the non-demolition
measurement of the quantum state [37,38]. That is to say, we can probe the
quantum information of the quantum state, but the state is not disturbed.
The paper is organized as follows. In Sec. II, we present a scheme to probe
the quantum entanglement, the quantum discord and the classical correlation,
but they are not disturbed in the probe procedure. In Sec. III, we give our
another model to demonstrate the non-demolition measurement of our given
quantum state. The conclusion is drawn finally.

\section{Probing concurrence, quantum and classical correlations without
disturbing them}

Suppose two qubits are prepared in the unknown one-parameter quantum state%
\begin{equation}
\rho _{0}=\left(
\begin{array}{cccc}
0 & 0 & 0 & 0 \\
0 & \frac{x}{2} & 1-\frac{3x}{2} & 0 \\
0 & 1-\frac{3x}{2} & \frac{x}{2} & 0 \\
0 & 0 & 0 & 1-x%
\end{array}%
\right) ,
\end{equation}%
where $x\in \lbrack \frac{1}{2},1]$. We introduce a third qubit as a probe
qubit to interact with the two given qubit in $\rho _{0}$. It will be shown
that one can read the entanglement, the quantum correlation and the
classical correlation of $\rho _{0}$ by measuring the third qubit, but the
entanglement, quantum correlation and the classical correlation after the
interaction are not disturbed.
\begin{figure}[tbp]
\includegraphics[width=2\columnwidth]{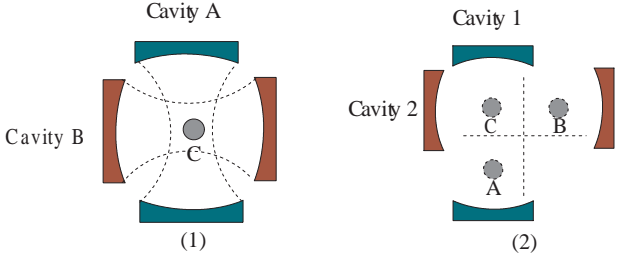}
\caption{(a) Schematic illustration of our probing quantum entanglement,
quantum discord and classical correlation. A two-level atom is surrounded by
two optical cavities. (b) Schematic illustration of non-demolition
measurement of quantum state. Two identical atoms are trapped in an optical
cavity, respectively. Another two-level atom as a probe qubit interacts with
the two optical cavity modes simultaneously.}
\label{1}
\end{figure}

Now we will discuss our scheme in the frame of cavity QED. Our system
includes one two-level atom labeled by C surrounded by two identical optical
cavities labeled by $A$ and $B,$see Fig. 1 (a). We suppose the initial state
of the two cavity modes is $\rho _{0}$ that will be detected and the atom C
serves as the probe qubit with an excited state $|e\rangle _{C}$ and a
ground state $|g\rangle _{C}$. The Hamiltonian of the joint system is given
by (setting $\hbar =1$)
\begin{equation}
H=\omega _{a}\sigma ^{+}\sigma ^{-}+\sum_{i=A,B}\omega _{c}a_{i}^{\dagger
}a_{i}+\frac{g}{\sqrt{2}}\sum_{i=A,B}(\sigma ^{+}a_{i}+a_{i}^{\dagger
}\sigma ^{-}),
\end{equation}%
where $\omega _{a}$ and $\omega _{c}$ are the frequencies of the atomic
transition and the cavity modes respectively; $\frac{g}{\sqrt{2}}$ is the
coupling constant of the atom and cavity mode and the $\sqrt{2}$ is for the
calculatable convenience; $\sigma ^{+}=|e\rangle \langle g|$ and $\sigma
^{-}=|g\rangle \langle e|$ are the raising and lowering operators of the
atom; $a_{i}$ is the annihilation operator of the $i$th cavity.

Under the resonant case, i.e. $\omega _{a}=\omega _{c}$, one can obtain the
Hamiltonian in the interaction picture as,
\begin{equation}
H_{I}=\frac{g}{\sqrt{2}}[\sigma ^{+}a_{A}+\sigma ^{-}a_{A}^{\dagger }+\sigma
^{+}a_{B}+\sigma ^{-}a_{B}^{\dagger }].
\end{equation}%
If the initial state of atom C is prepared in $|g\rangle _{C}$, the
evolution of the joint system of the two cavities and the atom can be given
by%
\begin{equation}
\rho (t)=\exp (-iH_{I}t)\left( \rho _{0}\otimes |g\rangle _{C}\left\langle
g\right\vert \right) \exp (iH_{I}t).
\end{equation}%
After a simple calculation, one can find that
\begin{equation}
\rho _{AB}(t)=\left(
\begin{array}{cccc}
\left( 1-x\right) \cos ^{2}(gt) & 0 & 0 & 0 \\
0 & \frac{x}{2} & \frac{2-3x}{2} & 0 \\
0 & \frac{2-3x}{2} & \frac{x}{2} & 0 \\
0 & 0 & 0 & \left( 1-x\right) \sin ^{2}(gt)%
\end{array}%
\right) ,
\end{equation}%
and
\begin{equation}
\rho _{C}(t)=\left(
\begin{array}{cc}
2\left( 1-x\right) \sin ^{2}(gt) & 0 \\
0 & 2\left( 1-x\right) \cos ^{2}(gt)+2x-1%
\end{array}%
\right) .
\end{equation}%
\

Next, we will employ Wootters' concurrence [2] as entanglement measure, the
quantum discord [9,10] as the measure of quantum correlation to discuss
their invariability after our detection. Concurrence of a bipartite quantum
state $\rho _{AB}$ is defined as
\begin{equation}
C(\rho _{AB})=\max \{0,\lambda _{1}-\lambda _{2}-\lambda _{3}-\lambda _{4}\},
\end{equation}%
where the $\lambda _{i}$ are the square roots of the eigenvalues of the
non-Hermitian matrix $\rho _{AB}\tilde{\rho}_{AB}$ with $\tilde{\rho}%
_{AB}=(\sigma _{y}\otimes \sigma _{y})\rho _{AB}^{\ast }(\sigma _{y}\otimes
\sigma _{y})$ in decreasing order. Substitute $\rho _{AB}(t)$ and $\rho _{0}$
into eq. (7), one can easily obtain
\begin{equation}
C(\rho _{AB}(t))=\max \{0,\left\vert 2-3x\right\vert -\left( 1-x\right)
\left\vert \sin (2gt)\right\vert \}
\end{equation}%
and%
\begin{equation}
C(\rho _{0})=\left\vert 2-3x\right\vert .
\end{equation}%
It is obvious that $C(\rho _{AB}(t_{n}))=C(\rho _{0})$ when $t_{n}=\frac{%
n\pi }{2g},n=0,1,2,\cdots $. At the same time, if one measures $\sigma
_{z}=\left(
\begin{array}{cc}
1 & 0 \\
0 & -1%
\end{array}%
\right) $ on $\rho _{C}(t_{n})$, one will get
\begin{equation}
C(\rho _{0})=\frac{\left\vert 3\left\langle \sigma _{z}(t_{n})\right\rangle
-1\right\vert }{4},
\end{equation}%
where $\left\langle \sigma _{z}(t_{n})\right\rangle =$Tr$\left[ \rho
_{C}(t_{n})\sigma _{z}\right] .$ That is to say, in the current ideal case
(there is no noise), so long as we measure $\sigma _{z}$ on $\rho
_{C}(t_{n}) $ at $t_{n}=\frac{n\pi }{2g}$, we can probe the concurrence of $%
\rho _{0}$ without disturbing the concurrence of $\rho _{AB}$. However, when
in the realistic scenario, one has to select $t_{1}=\frac{\pi }{2g}$, by
which one can reduce the time of interaction of the whole system with the
unavoidable noise.

Next, we will show that the quantum and the classical correlations can not
be disturbed by our detection. Quantum discord is used to measure the
quantum correlation between two subsystems. For a bipartite quantum system $%
\rho ^{ab}$ with $\rho ^{a}$ ($\rho ^{b}$) denoting the reduced density
matrix of subsystem $a$ ($b$), then the quantum discord between subsystems $%
a $ and $b$ can be defined as follows
\begin{equation}
\mathcal{Q}(\rho ^{ab})=\mathcal{I}(\rho ^{ab})-\mathcal{C}(\rho ^{ab}),
\end{equation}%
where
\begin{equation}
\mathcal{I}(\rho ^{ab})=S(\rho ^{a})+S(\rho ^{b})-S(\rho ^{ab})
\end{equation}%
is the quantum mutual information and $\mathcal{C}(\rho ^{ab})$ is the
classical correlation between the two subsystems. In particular, the
classical correlation is given by
\begin{equation}
\mathcal{C}(\rho )=\mathrm{max}_{\{B_{k}\}}[S(\rho ^{a})-S(\rho |\{B_{k}\})],
\end{equation}%
where $\{B_{k}\}$ is a set of von Neumann measurements performed on
subsystem $b$ locally, $S(\rho |\{B_{k}\})=\sum_{k}p_{k}S(\rho _{k})$ is the
quantum conditional entropy, $\rho _{k}=(\mathbb{I}\otimes B_{k})\rho (%
\mathbb{I}\otimes B_{k})/\mathrm{Tr}(\mathbb{I}\otimes B_{k})\rho (\mathbb{I}%
\otimes B_{k})$ is the conditional density operator corresponding to the
outcome labeled by $k$, and $p_{k}=\mathrm{Tr}(\mathbb{I}\otimes B_{k})\rho (%
\mathbb{I}\otimes B_{k})$. Here $\mathbb{I}$ is the identity operator
performed on subsystem $a$.

For the bipartite quantum state $\rho _{0}$ and $\rho _{AB}(t_{n})$, one can
analytically calculate the quantum and the classical correlations according
to Ref. [15]. In order to explicitly show the invariability of quantum
discord and classical correlation after our detection, we would like to
first consider the density matrix
\begin{equation}
\rho =\left(
\begin{array}{cccc}
\rho _{11} & 0 & 0 & 0 \\
0 & \rho _{22} & \rho _{23} & 0 \\
0 & \rho _{32} & \rho _{33} & 0 \\
0 & 0 & 0 & \rho _{44}%
\end{array}%
\right) ,
\end{equation}%
where all the entries are real. Based on eq. (12), one can get
\begin{equation}
\mathcal{I}(\rho _{AB}(t))=S(\rho _{AB}^{A})+S(\rho
_{AB}^{B})+\sum_{j=1}^{4}\lambda _{j}\log _{2}\lambda _{j},
\end{equation}%
where
\begin{eqnarray}
{S}(\rho _{AB}^{A}) &=&-[(\rho _{11}+\rho _{22})\log _{2}{(\rho _{11}+\rho
_{22})} \\
&&+(\rho _{33}+\rho _{44})\log _{2}{(\rho _{33}+\rho _{44})}],  \notag \\
S(\rho _{AB}^{B}) &=&-[(\rho _{11}+\rho _{33})\log _{2}{(\rho _{11}+\rho
_{33})} \\
&&+(\rho _{22}+\rho _{44})\log _{2}{(\rho _{22}+\rho _{44})}],  \notag
\end{eqnarray}%
and
\begin{equation}
\lambda _{1}=\rho _{11},\lambda _{2}=\rho _{22}-\rho _{23},\lambda _{3}=\rho
_{22}+\rho _{23},\lambda _{4}=\rho _{44}\newline
.
\end{equation}%
The classical correlation $\mathcal{C}(\rho _{AB}(t))$ can be given by
\begin{equation}
\mathcal{C}(\rho _{AB}(t))=S(\rho _{AB}^{A})-\mathtt{min}_{{Bi}}[S(\rho
_{AB}(t)|\{B_{i}\})],
\end{equation}%
where

\begin{equation}
\mathtt{min}_{{Bi}}[S(\rho _{AB}(t)|\{B_{i}\})]=\left\{
\begin{array}{cc}
(\rho _{22}+\rho _{33})\log _{2}(\rho _{22}+\rho _{33})-\rho _{22}\log
_{2}\rho _{22}-\rho _{33}\log _{2}\rho _{33}, & \rho _{44}\leq 0.4716 \\
1-\frac{1}{2}[(1-\theta )\log _{2}(1-\theta )+(1+\theta )\log _{2}(1+\theta
)], & \rho _{44}>0.4716%
\end{array}%
\right.
\end{equation}%
with $\theta =\sqrt{(\rho _{11}-\rho _{44})^{2}+4\rho _{23}{}^{2}}$.

From the above calculation, one can find that $\rho _{11}$ and $\rho _{44}$
are symmetric in eqs. (15-18,20) if $\rho _{22}=\rho _{33}$. In particular,
one can also find that the difference between $\rho _{0}$ and $\rho
_{AB}(t_{n})$ is the exchange of $\rho _{11}$ and $\rho _{44}$. That is to
say, $\rho _{0}$ and $\rho _{AB}(t_{n})$ have the same quantum discord and
the same classical correlation. What's more, since $\left\langle \sigma
_{z}(t_{n})\right\rangle =3-4x$ from eq. (10) and all the entries of $\rho
_{0}$ given in eq. (1) are the function of $x$, one can conclude that the
quantum discord $\mathcal{Q}(\rho _{0})$ and the classical correlation $%
\mathcal{C}(\rho _{0})$ can be obtained by measuring $\sigma _{z}$ on $\rho
_{C}(t_{n})$. Namely, we can detect the quantum and the classical
correlations of $\rho _{0}$ without disturbing them.

Finally, we would like to discuss the influence of the noise. As we know,
the decoherence will inevitably happen in $\rho _{0}$ if the two cavity
modes interact with noise no matter whether we introduce the third qubit to
probe the system. So we want to emphasize here that our detection has no
influence on the concurrence, the quantum discord and the classical
correlation of the system. From the viewpoint of noise, we say that the
noise due to our detection will have slight influence on the system. In this
sense, we only need to consider the noise relevant to the third
qubit-----the spontaneous emission of atom C.
\begin{figure}[tbp]
\includegraphics[width=1\columnwidth]{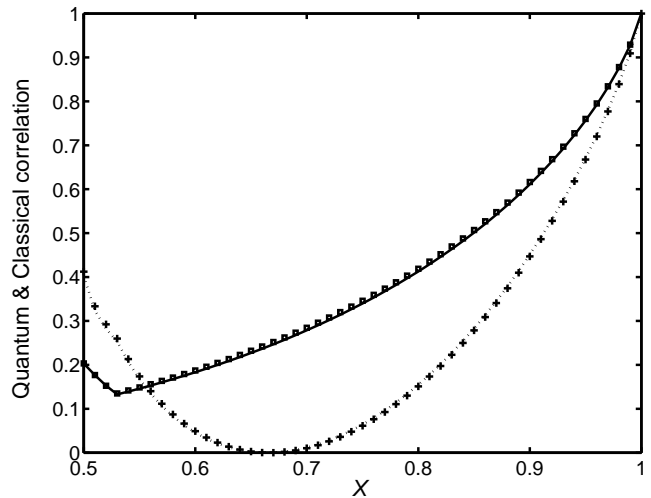}
\caption{(Dimensionless) Quantum discord and classical correlation vs. x.
The solid line and the square line correspond to the classical correlation
with and without spontaneous emission, respectively. The dotted line and the
plus line correspond to the quantum discord with and without spontaneous
emission, respectively. Here, $\protect\gamma =0.1g$. }
\label{1}
\end{figure}
In the interaction picture, the master equation governing the evolution of
the \textquotedblleft cavity+atom\textquotedblright\ system can be given by%
\begin{equation}
\dot{\rho}=-i[H_{I},\rho ]+\gamma (2\sigma ^{-}\rho \sigma ^{+}-\sigma
^{+}\sigma ^{-}\rho -\rho \sigma ^{+}\sigma ^{-}),
\end{equation}%
where $\gamma $ is the atomic spontaneous emission rate and $H_{I}$ is
defined as eq. (3). The initial state of the cavities is given by eq. (1)
and the atom is prepared in the ground state initially. We have numerically
solved the master equation. The concurrence $C(\rho _{AB}(t_{n}))$, the
quantum discord $\mathcal{Q}(\rho _{AB}(t_{n}))$ and the classical
correlation $\mathcal{C}(\rho _{AB}(t_{n}))$ with and without noise vs. $x$
are plotted in Fig. 2 and $\left\langle \sigma _{z}(t_{n})\right\rangle $
with and without noise vs. $x$ are plotted in Fig. 3. We set $\gamma =0.1g$.
From the figure, one can find that the spontaneous emission has slight
influence on the entanglement and correlation of $\rho _{0}$ as well as our
detection. In particular, the quantum discord of the original state is
influenced more slightly by the small $\gamma $. In addition, we also find a
very interesting phenomenon that the existence of $\gamma $ might lead to
the increment of quantum discord for some large $x$, such as $x>0.8$.
\begin{figure}[tbp]
\includegraphics[width=1\columnwidth]{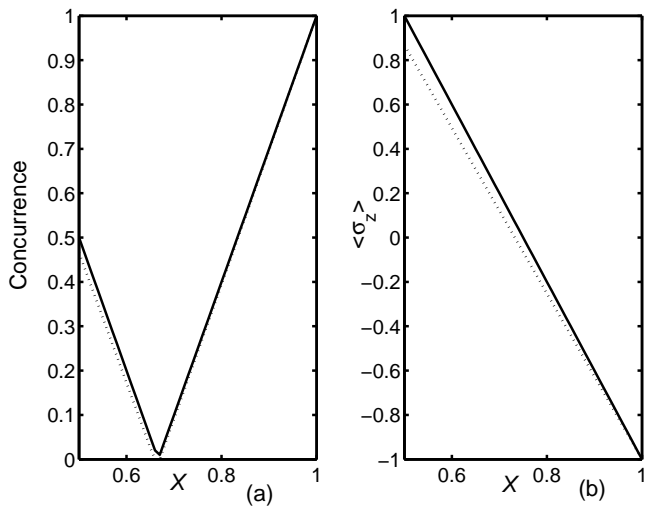}
\caption{(Dimensionless)(a) Concurrence vs. $x$. (b) $<\protect\sigma _{z}>$
vs. $x$. The solid line and the dotted line in both the figures correspond
to the case with and without spontaneous emission, respectively. Here, $%
\protect\gamma =0.1g$.}
\label{1}
\end{figure}

\section{\protect\bigskip Non-demolition measurement of the quantum state}

In the previous section, we have presented a model to probe the concurrence
and correlation of a type one-parameter quantum state without disturbing
them. In fact, in a different framework, one can realize the non-demolition
measurement of the quantum state. In this case, we consider two identical
two-level atoms A and B which are trapped in an optical cavity,
respectively. In particular, the two cavities are arranged to be crossed as
is sketched in Fig. 1 (b). We will introduce a third atom C as a probe
qubit. One can find that the measurement on the qubit C can reveal the
information of the quantum state of the joint system `A+B' without
disturbing it.

The Hamiltonian of our whole system is given by (setting $\hbar =1$)
\begin{eqnarray}
H &=&\sum_{i=A,B,C}\omega _{i}\sigma _{i}^{+}\sigma _{i}^{-}+\sum \upsilon
a_{j}^{\dagger }a_{j}+\frac{g}{\sqrt{2}}\sum_{j=1,2}(\sigma
_{C}^{+}a_{j}+a_{j}^{\dagger }\sigma _{C}^{-})  \notag \\
&&+\frac{g}{\sqrt{2}}(\sigma _{A}^{+}a_{1}+a_{1}^{\dagger }\sigma _{A}^{-})+%
\frac{g}{\sqrt{2}}(\sigma _{B}^{+}a_{2}+a_{2}^{\dagger }\sigma _{B}^{-}),
\end{eqnarray}%
where $\omega _{i},i=A,B,C$ and $\omega _{c}$ are the frequencies of the
atomic transition and the cavity modes respectively; $\frac{g}{\sqrt{2}}$ is
the coupling constant of the atom and cavity mode; $\sigma _{i},i=A,B,C$
denotes the lowering operators of the atom $i$; $a_{j}$ is the annihilation
operator of the $j$th cavity. In addition, we let $\omega _{A}=\omega _{B}$
and $\omega _{C}-\omega _{A}=g^{2}/2\delta $ with $\delta =\omega
_{C}-\upsilon $. If there is not any photons initially in both the cavities,
under the large detuning condition, i.e. $\delta \gg \frac{g}{\sqrt{2}}$,
one can obtain the effective Hamiltonian of the joint system in a proper
rotation frame as
\begin{equation}
H_{eff}=\frac{g^{2}}{2\delta }(\sigma _{A}^{-}\sigma _{C}^{+}+\sigma
_{A}^{+}\sigma _{C}^{-}+\sigma _{B}^{-}\sigma _{C}^{+}+\sigma _{B}^{+}\sigma
_{C}^{-}).
\end{equation}%
Thus, the evolution of the system can be given by the time evolution
operator
\begin{equation}
U(t)=\exp (-iH_{eff}t).
\end{equation}%
Now we turn to the original state $\rho _{0}$ given in eq. (1). After the
evolution governed by $U(t)$, one can find that
\begin{equation}
\rho _{0}\overset{U(t_{m}):|e\rangle _{C}\langle e|}{\rightarrow }\rho
_{AB}(t_{m})=(\sigma _{x}\otimes \sigma _{x})\rho _{0}(\sigma _{x}\otimes
\sigma _{x}),
\end{equation}%
since one can find that $H_{eff}$ have the consistent form with the
Hamiltonian of eq. (3). Here $|e\rangle _{C}\langle e|$ means that the
initial state of the probe atom is $|e\rangle _{C}$ and $t_{m}=\frac{\delta
m\pi }{g^{2}}(m=1,2,3,...)$. The quantum state of atom C at this time is
consistent with that in eq. (6). It is very interesting that one can find
\begin{equation}
\rho _{AB}(t_{m})\overset{U(\tau _{n}):|g\rangle _{C}\langle g|}{\rightarrow
}\rho _{AB}(\tau _{n})=\rho _{0},
\end{equation}%
which means the initial state $\rho _{AB}(t_{m})$ evolves to $\rho
_{AB}(\tau _{n})$ with the initial atom C in $\left\vert g\right\rangle _{C}$
and the evolution time $\tau _{n}=\frac{\delta n\pi }{g^{2}}(n=1,2,3,...)$.
In addition, the state of atom C at $\tau _{n}$ are $\sigma _{x}\rho
_{C}(t_{m})\sigma _{x}$. So when we probe the quantum state $\rho _{0}$, we
can first select the initial atom C in $|e\rangle _{C}$ and after the system
evolve for $t_{m}$, we remove the atom C, and then select another atom C' in
$|g\rangle _{C}$ to interact with the residual atoms A and B for $\tau _{n}$%
. In this way, the final state of atoms A and B will stay in the initial
state $\rho _{0}$. One can repeat the same procedure on the atoms A and B to
complete the quantum measurement statistics of atom C. But from our
calculation, one will find that $\rho _{0}$ is not disturbed. That is to
say, one has completed the non-demolition measurement of the quantum state.

Besides the non-demolition measurement of the quantum state, we would like
to compare this model with that in Sec. II. One can find that the intial
state of the probe qubit must be $|g\rangle _{C}$ in the model of Sec. II,
otherwise the original state of $\rho _{0}$ will have to be disturbed. That
is to say, in Sec. II, the quantum state can only be probed once. The
residual quantum state has the same entanglement, discord and classical
correlation as $\rho _{0}$. However, in the model of this section, one can
alternately select $|g\rangle _{C}$ and $|e\rangle _{C}$ as the initial
quantum state of the probe qubit. One divides the measurement outcomes into
two groups according to the different choice of $|g\rangle _{C}$ and $%
|e\rangle _{C}$. For each group (or any one of the two groups), one can
complete the measurement statistics. In this way, we can implement the
measurement of entanglement and correlation without disturbing them.

\section{Conclusion and discussion}

We have presented a scheme for a type of one-parameter quantum state to
probe quantum entanglement, quantum discord and classical correlation
without disturbing them. We also analyze how the spontaneous emission of our
probe qubit influences our detection, which shows that small spontaneous
emission rate of the probe atom has slight influence on the detection. In
particular, one can find that the influence on the quantum discord and
classical correlations is much weaker than the concurrence. It is very
interesting that we have found that spontaneous emisstion might benefit to
the quantum discord for some quantum states. In addition, we also present a
scheme to implement the non-demolition measurement of the quantum state.
That is to say, based on our scheme, one can extract the information of the
quantum state, but the quantum state per se is not disturbed. However, it
seems that our scheme is not universal for a general quantum state. How to
develop a scheme for multiple-parameter quantum state to probe the
concurrence, quantum discord or classical correlation, even the quantum
state without disturbing them deserves our forthcoming efforts.

\section{Acknowledgement}

This work was supported by the National Natural Science Foundation of China,
under Grant No. 10805007 and No. 10875020, and the Doctoral Startup
Foundation of Liaoning Province.

\end{document}